\documentclass[
    ,final            
  ]
  {aipproc}

\layoutstyle{6x9}

\begin{document}

\title{Dynamical properties of a family of collisionless models of elliptical galaxies}

\author{G. Bertin}{
  address={Dipartimento di Fisica, Universit\`{a} di Milano, via
Celoria 16, I-20133 Milano, Italy}
}

\author{M. Trenti}{
  address={Scuola Normale Superiore, piazza dei Cavalieri 7, I-56126 Pisa, Italy}
}

\begin{abstract}
N-body simulations of collisionless collapse have offered
important clues to the construction of realistic stellar dynamical
models of elliptical galaxies. Such simulations confirm and
quantify the qualitative expectation that rapid collapse of a
self-gravitating collisionless system, initially cool and
significantly far from equilibrium, leads to incomplete
relaxation, that is to a quasi-equilibrium configuration
characterized by isotropic, quasi-Maxwellian distribution of
stellar orbits in the inner regions and by radially biased
anisotropic pressure in the outer parts. In earlier studies, as
illustrated in a number of papers several years ago (see
\cite{ber93} and references therein), the attention was largely
focused on the successful comparison between the models
(constructed under the qualitative clues offered by the N-body
simulations mentioned above) and the observations.  In this paper
we revisit the problem of incomplete violent relaxation, by making
a direct comparison between the detailed properties of a family of
distribution functions and those of the products of collisionless
collapse found in N-body simulations.

\end{abstract}

\maketitle

\section{Introduction}

We examine the results of a set of simulations of collisionless
collapse of a stellar system and focus on the fit of the final
quasi-equilibrium configurations by means of the so-called
$f^{(\nu)}$ models (see \cite{sti87}, \cite{ber03}). The models,
constructed on the basis of theoretical arguments suggested by
statistical mechanics, are associated with realistic luminosity
and kinematic properties, suitable to represent elliptical
galaxies. All the simulations have been performed using a new
version \cite{tren04} of the original code of van Albada (see
\cite{van82}; the code evolves a system of simulation particles
interacting with one another via a mean field calculated from a
spherical harmonic expansion of a smooth density distribution),
starting from various initial conditions characterized by
approximate spherical symmetry and by a small value of the virial
parameter $u$ ($u = -2K/W < 0.2$; here $K$ and $W$
represent total kinetic and gravitational energy). From such
initial conditions, the collisionless ``gravitational plasma"
evolves undergoing incomplete violent relaxation \cite{lyn67}.

The models, in spite of their simplicity and of their spherical
symmetry, turn out to describe well and in detail the end products
of our simulations. We should note that the present formulation
neglects the presence of dark matter. Since there is convincing
evidence for the presence of dark matter halos also in elliptical
galaxies, this study should be extended to the case of
two-component systems, before a satisfactory comparison with
observed galaxies can eventually be claimed.

\section{N-body experiments}  \label{sec:init}

Most simulations have been run with $8 \times 10^5$ particles; the
results have been checked against varying the number of particles.
The units chosen are $10^{11}~M_{\odot}$ for mass, $10~kpc$ for
length, and $10^8~yr$ for time. In these units the gravitational
constant is $G \approx 4.497$, the mass of the model is $2$, and
the initial radius of the system $1$.

We focus on clumpy initial conditions. These turn out to lead to
more realistic final configurations (\cite{van82}, \cite{lon91}).
They can be interpreted as a way to simulate the merging of
several smaller structures to form a galaxy. The clumps are
uniformly distributed in space, within an approximate spherical
symmetry; the ellipticity at the beginning of the simulation is
small, so that the corresponding initial shape would resemble that
of $E2-E3$ galaxies. We performed several runs varying the number
of clumps and the virial ratio $u$ in the range from $0.05$ to
$0.2$. Usually the clumps are cold, i.e. their kinetic energy is
all associated with the motion of their center of mass.

The simulations are run up to time $t = 10$, when the system has
settled into a quasi-equilibrium. The violent collapse phase takes
place within $t \approx 1$ . The final configurations are
quasi-spherical, with shapes that resemble those of $E2-E3$
galaxies. The final equilibrium half-mass radius $r_M$ is
basically independent of the value of $u$. The central
concentration achieved is a function of the initial virial
parameter $u$, as can be inferred from the conservation of maximum
density in phase space \cite{lon91}.

\section{Physically justified models} \label{sec:fv}

For a spherically symmetric kinetic system, extremizing the
Boltzmann entropy at fixed values of the total mass, of the total
energy, and of an additional third quantity $Q$, defined as $ Q =
\int J^{\nu} |E|^{-3 \nu/4} f d^3q d^3p$, leads (\cite{sti87}; see
also \cite{ber03}) to the distribution function $f^{(\nu)} = A
\exp {[- a E - d (J^2/|E|^{3/2})^{\nu/2}]}$, where $a$, $A$, $d$,
and $\nu$ are positive real constants; here $E$ and $J$ denote
single-star specific energy and angular momentum. At fixed value
of $\nu$, one may think of these constants as providing two
dimensional scales (for example, $M$ and $Q$) and one
dimensionless parameter, such as $\gamma = ad^{2/\nu}/(4 \pi GA)$.
In the following we will focus on values of $\nu$ ranging from 3/8
to 1 .

The corresponding models are constructed by solving the Poisson
equation for the self-consistent mean potential $\Phi(r)$
generated by the density distribution associated with $f^{(\nu)}$.
At fixed value of $\nu$, the models thus generated make a
one-parameter family of equilibria, described by the concentration
parameter $\Psi = -a \Phi(0)$, the dimensionless depth of the
central potential well.

The projected density profile of the models is very well fitted by
the $R^{1/n}$ law (for a definition of the law and of the
effective radius $R_e$, see \cite{ber02}), with the index $n$
varying in the range from $2$ to $6$, depending on the precise
value of $\nu$ and on the concentration parameter $\Psi$. At
worst, the residuals from the $R^{1/n}$ law are less than $0.25$
mag over a range of more than $8$ mag, while in general the
residuals are less than $0.1$ mag in a range of more than $10$
mag.

The models are isotropic in the inner regions, while they are
characterized by almost radial orbits in the outer parts. The
local value of pressure anisotropy can be measured by $\alpha(r) =
2 - (\langle p^2_{\theta}\rangle + \langle p^2_{\phi}
\rangle)/\langle p^2_r \rangle$. The form of the transition from
isotropy ($\alpha \approx 0$) to radial anisotropy ($\alpha
\approx 2$) is governed by the index $\nu$: higher values of $\nu$
give rise to a sharper transition. The anisotropy radius, that is
the location where the transition takes place (with $\alpha = 1$),
is close to the half-mass radius of the models \cite{ber03}.

\section{Fits and phase-space properties} \label{sec:fit}

In order to study the output of the simulations we compare the
density and the anisotropy profiles, $\rho (r)$ and $\alpha (r)$,
of the end products with the theoretical profiles of the
$f^{(\nu)}$ family of models; smooth simulation profiles are
obtained by averaging over time, based on a total of $20$
snapshots taken from $t = 8$ to $t = 10$. For the fitting models,
the parameter space explored is that of an equally spaced grid in
$(\nu,\Psi)$, with a subdivision of $1/8$ in $\nu$, from $3/8$ to
$1$, and of $0.2$ in $\Psi$, from $0.2$ to $14.0$; the mass and
the half-mass radius of the models are fixed by the scales set by
the simulations. A least $\chi^2$ analysis is performed, with the
error bars estimated from the variance in the time average process
used to obtain the smooth simulation profiles. A critical step in
this fitting procedure is the choice of the relative weights for
the density and the pressure anisotropy profiles. We adopted equal
weights for the two terms, checking a posteriori that their
contributions to $\chi^2$ are of the same order of magnitude.

\begin{figure}
\includegraphics[height=.31\textheight]{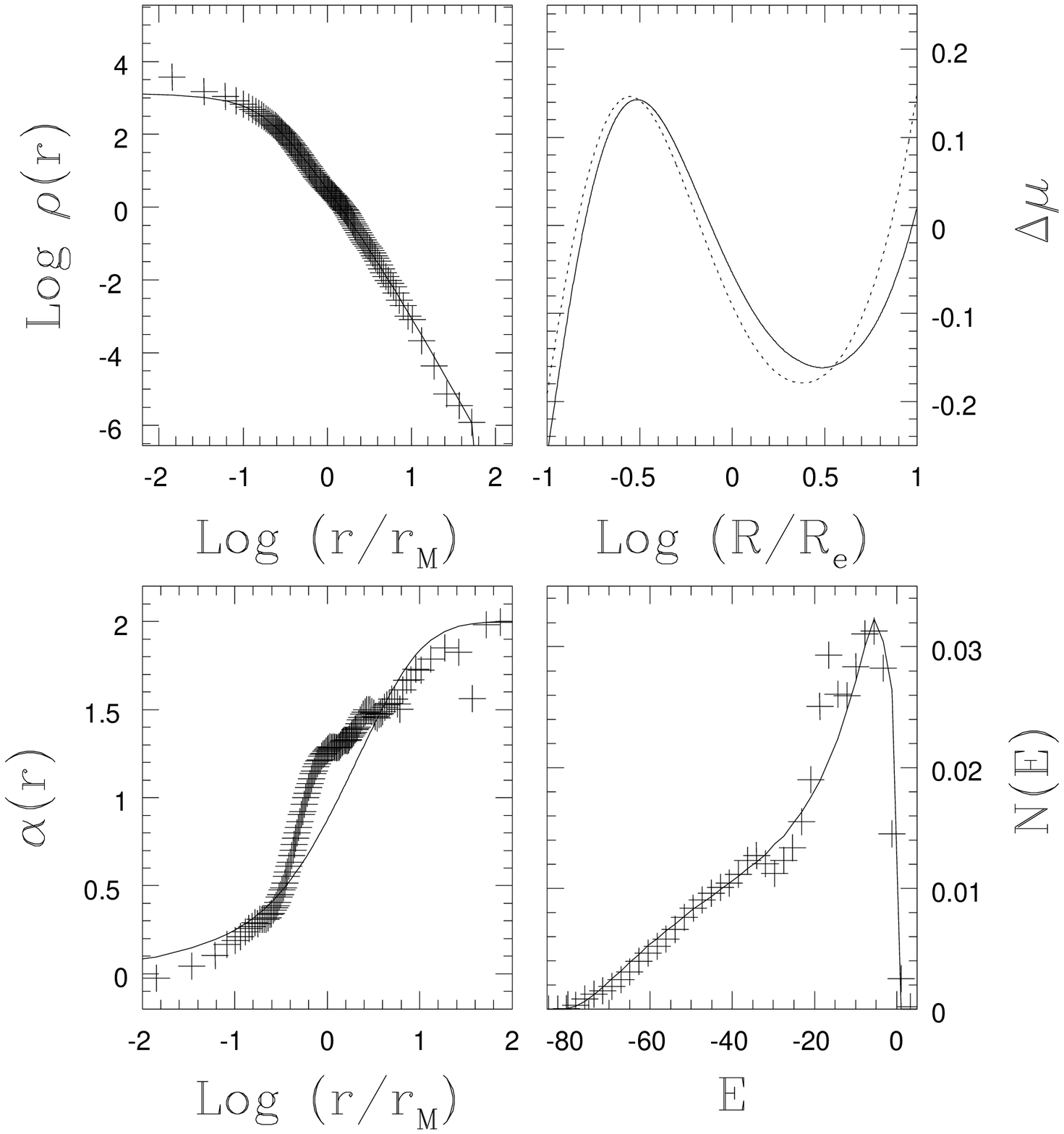}
\includegraphics[height=.3\textheight]{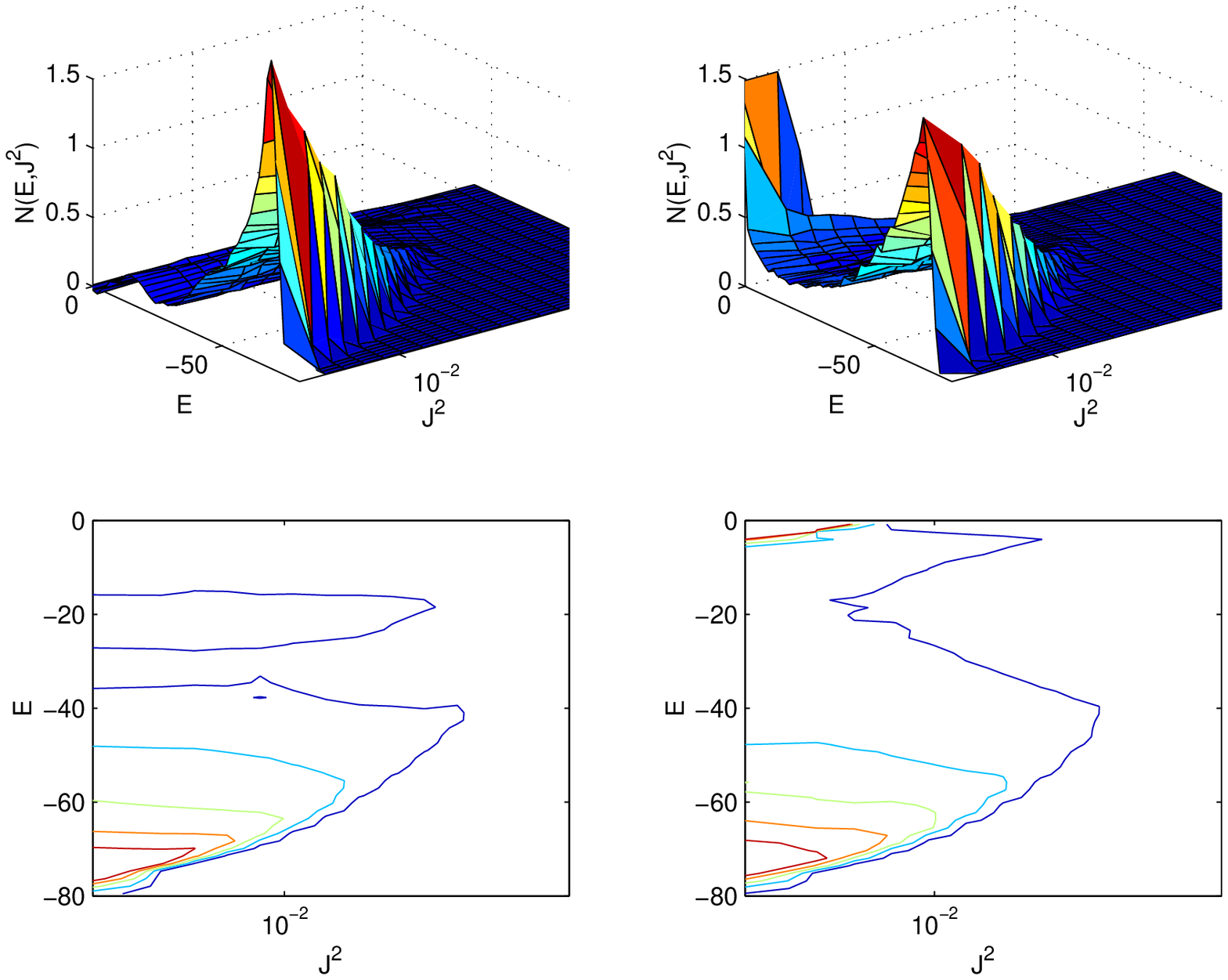}
\caption{Comparison between the results of one simulation with
initial $u = 0.15$, called $C2$ (starting from 20 cold clumps of
radius $0.4$), and the best fit $f^{(\nu)}$ model ($\nu = 0.5$,
$\Psi = 6.2$). {\it Left set of four panels}: Density as measured
from the simulation (crosses) vs. the best fit profile (top left).
Residuals (in magnitudes) from the $R^{1/4}$ (dotted line) and
from the $R^{1/n}$ (with $n=4.46$) law for the best fit
projected-density profile (top right). Anisotropy profile of the
simulation (crosses) vs. the best fit profile (bottom left).
Energy density distribution $N(E)$ (bottom right). {\it Right set
of four panels}: Final phase space density $N(E,J^2)$ (left
column), compared with that of the best fitting $f^{(\nu)}$ model
(right column). The model curve for $N(E)$ and surface for
$N(E,J^2)$ have been computed by a Monte Carlo sampling of phase
space. \label{fig:simC2}}

\end{figure}

As exemplified by Fig.~\ref{fig:simC2}, the density of all the
simulations is well represented by the best fit $f^{(\nu)}$
profile over the entire radial range. The fit is satisfactory not
only in the outer parts, where the density falls under a threshold
value that is {\it nine orders of magnitude} smaller than that of
the central density, but also in the inner regions, where, in
principle, there could be problems of numerical resolution. We
have performed simulations with different numbers of particles,
without noticing significant changes in the relevant profiles and
in the quality of the fits. The successful comparison between
models and simulations is also interesting because, depending on
the initial virial parameter $u$, the less concentrated end
products are fitted by a density profile which, if projected along
the line-of-sight, exhibits different values of $n$ in an
$R^{1/n}$ best fit analysis. In turn, this may be interpreted in
the framework of the proposed weak homology of elliptical galaxies
\cite{ber02}.

To some extent, the final anisotropy profiles for clumpy initial
conditions are found to be sensitive to the detailed choice of
initialization. In other words,  runs starting from initial
conditions with the same parameters, but with a different
\emph{seed} in the random number generator, give rise to slightly
different profiles. The agreement between the simulation and the
model profiles is good (see Fig.~\ref{fig:simC2}), except for
about $20 \%$ ``irregular" cases; the origin of these is not
clear, and we argue that they correspond to inefficient mixing in
phase space during collapse (a discussion of these issues will be
provided in a separate paper, currently in preparation).

At the level of phase space, we have performed two types of
comparison, one involving the energy density distribution $N(E)$
and the other based on $N(E,J^2)$. The chosen normalization
factors are such that: $M = \int N(E) dE = \int N(E,J^2) dE dJ^2.$
In Fig.~\ref{fig:simC2} we plot the final energy density
distribution for a simulation run called $C2$ (a regular case)
with respect to the predictions of the best fit model identified
from the study of the density and pressure anisotropy
distributions. The agreement is very good, especially for the
strongly bound particles. In some simulations, the distribution of
less bound particles shows some deviations from the theoretical
expectations, especially in the ``irregular" cases, when the final
$N(E)$ exhibits a double peak; one peak is around $E = 0$, as
expected, while the other is located at the point where there was
a peak in $N(E)$ at the initial time $t=0$. We interpret this
feature as a signature of inefficient phase mixing. Finally, at
the deeper level of $N(E,J^2)$, simulations and models also agree
very well, as illustrated in the right set of four panels of
Fig.~\ref{fig:simC2}. For the case shown, the distribution contour
lines essentially coincide in the range from $E_{min}$ to $E
\approx - 30$; however, the theoretical model shows a peak located
near the origin, corresponding to an excess of weakly bound stars
on almost radial orbits.

\end{document}